\documentclass[revtex]{emulateapj}
\journalinfo{Accepted for publication in The Astrophysical Journal}

\newcommand{\HI}{{\sc H\,i}}
\newcommand{\HII}{{\sc H\,ii}}
\newcommand{\kms}{\rm km\ s^{-1}}
\newcommand{\radm}{\rm rad\ m^{-2}}
\newcommand{\radmsq}{\rm rad^2\ m^{-4}}

\shorttitle{Structure in the Rotation Measure Sky}
\shortauthors{Stil et.~al.}

\begin{document}

\title{Structure in the Rotation Measure Sky}

\author{J. M. Stil}
\author{A. R. Taylor}
\author{C. Sunstrum}
\affil{Institute for Space Imaging Science and Department of Physics and Astronomy, University of Calgary}

\begin{abstract}
An analysis of structure in rotation measure (RM) across the sky based
on the RM catalog of \citet{taylor2009} is presented.  Several
resolved RM structures are identified with structure in the local ISM,
including radio loops I, II, and III, the Gum nebula, and the
Orion-Eridanus super bubble. Structure functions (SFs) of RM are
presented for selected areas, and maps of SF amplitude and slope
across the sky are compared with H$\alpha$ intensity and diffuse
polarized intensity.  RM variance on an angular scale of $1\degr$ is
correlated with length of the line of sight through the Galaxy, with a
contribution from local structures. The slope of the SFs is less
concentrated to the Galactic plane and less correlated with length of
the line of sight through the Galaxy, suggesting a more local origin
for RM structure on angular scales $\sim 10\degr$. The RM variance is
a factor $\sim 2$ higher towards the SGP than towards the NGP,
reflecting a more wide-spread asymmetry between the northern and
southern Galactic hemispheres.  Depolarization of diffuse Galactic
synchrotron emission at latitudes $< 30\degr$ can be explained largely
by Faraday dispersion related to small-scale variance in RM, but the
errors allow a significant contribution from differential Faraday
rotation along the line of sight.
\end{abstract}


\section{Introduction}

The magnetic field of the Milky Way Galaxy is the only galactic
magnetic field that can be studied over $\sim$ 5 orders of magnitude,
on length scales less than a parcsec to several kiloparsec.  The
structure of this magnetic field is revealed by different observing
techniques that are usually sensitive to either the line-of-sight
component of the magnetic field (Faraday rotation, Zeeman splitting),
or the component of the magnetic field perpendicular to the line of
sight (stellar polarization, polarization of radio synchrotron
emission). Zeeman splitting provides a direct in-situ measurement of
the strength of the magnetic field, while Faraday rotation and stellar
polarization are propagation effects that require additional
information on the electron density or the extinction column of dust
along the line of sight. Galactic synchrotron emission at frequencies
of a few GHz suffers from strong depolarization from Faraday rotation
in the Galactic disk \citep[e.g.][]{sokoloff1998,gaensler2001,uyaniker2003}. Even at high frequencies
\citep[e.g. 20 GHz WMAP;][]{page2007} where Faraday rotation is not
significant, the summation of polarized emission with different
polarization angles over a long line of sight can be a significant
source of depolarization.

The large-scale structure of the Galactic magnetic field has been
revealed mostly by Faraday rotation of polarized emission from pulsars
and distant active galactic nuclei (AGN)
\citep[][among others]{simard1979,simard1980,han2006,brown2001,brown2003,brown2007,vallee2008,men2008,taylor2009}. The
angle change $\Delta\theta$ from Faraday rotation of polarized
emission of a compact radio source with negligible internal Faraday
rotation is given by
\begin{equation}
\Delta\theta = \lambda^2  0.812 \int_{\rm source}^{\rm observer} n_e {\bf{B}} \cdot d{\bf{l}} = \lambda^2 RM
\label{FR-eq}
\end{equation}
where $\lambda$ if the observing wavelength, $n_e$ is the electron
density, $\bf B$ is the magnetic field vector, and $d{\bf l}$ is a
line of sight segment pointing from the source to the observer so that
$RM$ has a positive sign if there is a net magnetic field component
along the line of sight in the direction of the observer. The angle
change $\Delta\theta$ can be determined from observations of the
polarization angle of a radio source as a function of wavelength.

The electron density along the line of sight must be determined
independently to derive the strength of the magnetic field.  For the
large-scale magnetic field of the Galaxy, a model of the distribution
of the electron density $n_e$ along the line of sight has been derived
from pulsar dispersion measures \citep{cordes2002}.  However, for
smaller length scales, an electron density distribution must be
proposed as an assumption that is largely unconstrained by current
observations, as significant angle changes can occur in an ionized
region that cannot be detected by its thermal emission
\citep{uyaniker2003}. Correlation between fluctuations in electron
density and magnetic field strength can bias the RM
upward or downward, depending whether there is a positive correlation
or an anti-correlation respectively between thermal electron density
and field strength \citep{beck2003}.

Structure in the Galactic magnetic field on smaller scales has been
apparent in polarization data of stars within 1 kpc \citep[e.g.][and
references therein]{axon1976,
mathewson1970,leroy1999,heiles2000,berdyugin2002, berdyugin2004}, and
polarization of diffuse Galactic synchrotron emission
\citep[e.g.][Landecker et al., in
prep.]{duncan1997,duncan1999,uyaniker1999,gaensler2001,wolleben2006,page2007,schnitzeler2007,wolleben2007,gao2010}. Diffuse
polarized emission has the advantage that it provides continuous
sampling of a foreground Faraday screen, but the signal is usually a
complicated integral of synchrotron emissivity and Faraday rotation
over the volume probed by the beam
\citep{sokoloff1998}. Equation~\ref{FR-eq} does not apply in this
case, but imaging polarized intensity and polarization angle, and RM
synthesis \citep{burn1966,brentjens2005} have revealed structures in
diffuse polarized emission
\citep[e.g.][]{gray1998,gaensler2001,uyaniker2003,schnitzeler2007,debruyn2009}.

RM structure functions (SFs)
\citep{simonetti1984,simonetti1986,minter1996} have been used to
obtain a statistical description of RM variance as a function of
angular scale. At high Galactic latitudes the variance in RM was found
to be independent of angular scale \citep{simonetti1984,sun2004}. At
lower latitudes, RM variance is often observed to increase with
angular scale \citep{simonetti1986,simonetti1992,sun2004}, sometimes
with a break around an angular scale of $1\degr$ above which the
variance remains constant or increases at a slower rate
\citep{haverkorn2003,haverkorn2006,haverkorn2007,haverkorn2008,roy2008}.
The break in the RM SF is usually interpreted as an outer scale in the
turbulence that gives rise to the RM variance. The smallest scale that
can be probed depends on the density of RM measurements, and angular
resolution of the radio data
\citep{simonetti1986,leahy1987}. \citet{gaensler2005} and
\citet{haverkorn2008} used depolarization of point sources as a
measure of RM variations on the angular scale of extragalactic radio
sources.

A database of 37,543 RMs that covers 80\% of the sky
\citep{taylor2009} was constructed recently from the NRAO VLA Sky
Survey \citep[NVSS; ][]{condon1998}, using the two frequency bands of
the NVSS separated by 70 MHz (center to center). The NVSS RMs are
reliable to $|RM| < 520\ \radm$.  Higher RMs are subject to
ambiguities because it was not possible to derive a unique RM using
the available constraints and noise in the data. This occurs mainly at
low Galactic latitude ($|b| \lesssim 5\degr$).

\section{Large-scale RM structures and the local ISM}

\begin{figure*}
%\resizebox{\textwidth}{!}{\includegraphics[angle=0]{f1.eps}}
\caption{ {\bf Figure available as separate jpg file} Correlation
between RM and H$\alpha$ intensity from
\citet{finkbeiner2003}. Positive RM are indicated by red circles,
while negative RMs are indicated by blue circles. Circles are scaled
according to RM amplitude.
\label{Ha_RM-fig}
}  
\end{figure*}

\begin{figure*}
%\resizebox{\textwidth}{!}{\includegraphics[angle=0]{f2.eps}}
\caption{ {\bf Figure available as separate jpg file}
Correlation between RM and polarized intensity
from \citet{wolleben2006}. See also \citet{wolleben2007} for a
discussion of the large-scale features in this image.
\label{DRAO_RM-fig}
}  
\end{figure*}

\citet{taylor2009} showed the signature of the large-scale magnetic
field in the solar neighborhood in these data including a reversal of
the sign of RM across the Galactic plane consistent with a quadrupole
magnetic field geometry. Assuming a symmetric electron density, they
found that the magnetic field strength south of the Galactic plane is
approximately two times the magnetic field strength in the
north. Towards the Galactic poles, \citet{taylor2009} found a positive
RM in the north and in the south, also stronger in the south.
\citet{map2010} found a significant mean RM in the region around the
South Galactic Pole (GPP), but not around the North Galactic Pole
(NGP).

Here we discuss RM structures on angular scales $\lesssim 90\degr$ and
their relation the local interstellar medium. Figures~\ref{Ha_RM-fig},
\ref{DRAO_RM-fig}, and \ref{ROSAT_RM-fig} show the RM database as
circles on a background of H$\alpha$ intensity \citep{finkbeiner2003},
diffuse polarized emission at 1.4 GHz \citep{wolleben2006}, and
diffuse X-ray emission \citep{snowden1998},
respectively. Figure~\ref{conditional_RM-fig} shows a different
representation of the RM sky towards the anticenter, excluding data
with $|RM| < 25\ \radm$, and showing all circles at the same size.
This representation emphasizes RM structure with amplitudes of a few
tens $\radm$, but also boundaries where the sign of RM changes.

Three large regions previously identified by \citet{simard1980} as
regions A, B, and C are responsible for most of the structure on
angular scales $\gtrsim 30\degr$. These structures are also
represented in other all-sky RM images with resolution $\sim 20\degr$
\citep{frick2001,JH2004,dineen2005,xu2006}. The denser sampling of the
present data reveals structure in more detail, and we adopt somewhat
different boundaries for regions A and C of \citet{simard1980} as
discussed below. The boundaries adopted here are listed in
Table~\ref{SFregions-tab}, and we will refer to region A$'$ and region
C$'$ instead. Region B in \citet{simard1980} refers to the Gum nebula,
and will be discussed as such.

Region A is a large area of negative RM for which \citet{simard1980}
listed the boundaries as $60\degr < l < 140\degr$, $-40\degr < b <
10\degr$.  With the superior sampling of the new data, we see
substantial substructure in this region. The most negative RMs occur
at intermediate latitudes in the region $80\degr < l < 150\degr$,
$-40\degr < b < -20\degr$. Distinctly smaller RM amplitudes are found
in the area $100\degr < l < 150\degr$, $-20\degr < b < -5\degr$,
except for an area with higher RM amplitudes in the longitude range
$120\degr < l < 130\degr$ (See also
Figure~\ref{conditional_RM-fig}). The mean RM in the area $100\degr <
l < 120\degr$, $-40\degr < b < -20\degr$ is $-67.7\ \radm$, with a
standard deviation of $38.5\ \radm$. The mean H$\alpha$ intensity in
this area is 1.8 R. The mean RM in the area $100\degr < l < 120\degr$,
$-20\degr < b < -5\degr$ is $-32.8\ \radm$ with a standard deviation
of $31.9\ \radm$. The mean background H$\alpha$ intensity in this
area, avoiding bright \HII\ regions is 5 R.  The section of region A
with higher RM amplitude is therefore separated from the Galactic
plane by a region with RM amplitudes a factor $\sim 2$ lower than
those at more negative latitudes, even though the H$\alpha$ intensity
is a factor $\sim 2.5$ higher than at more negative
latitudes. \citet{simard1980} suggested that region A extends across
the Galactic plane, to $b = +10\degr$, but this appears to be based on
only two sources in their Figure 1, and there is no indication of an
extension to positive latitudes in their Figure~2.  The data of
\citet{simard1980} sampled the intermediate region with low RM
amplitudes very sparsely.

Region A is bordered on the high-longitude side by a line where RM
nearly vanishes, and then reverses sign to become positive
\citep[e.g.][]{xu2006}. The boundary with near-zero RM runs
approximately at an angle of $45\degr$ inclined to the Galactic
equator from $(l,b) \approx (145\degr, -10\degr)$ to $(l,b) \approx
(170\degr, -35\degr)$ (Figure~\ref{conditional_RM-fig}), consistent
with the boundary of Radio Loop II \citep{berkhuijsen1973}. A spur of
bright polarized emission is visible in Figure~\ref{DRAO_RM-fig}
inside the region with negative RM, while strong depolarization is
found at the boundary where RM changes sign.  The H$\alpha$ image
shows $\sim 20\degr$ long filaments with a similar orientation with
respect to the Galactic equator, approximately coincidental with the
line where RM changes sign.

Region B of \citet{simard1980}, centered near $l \approx 255\degr$, $b
\approx 0\degr$, was identified with the Gum nebula by these
authors. Structure in RM amplitude has since been noticed in this area
in the form of a magnetized shell \citep{vallee1983,stil2007} that is
also visible in Figure~\ref{Ha_RM-fig}. We note also that there is no
clear reversal of the sign of RM across the Galactic equator in the
area of the Gum nebula.

Region C of \citet{simard1980} is an area with positive RM bounded by
$0\degr < l < 60\degr$, $0\degr < b < 60\degr$ and was noticed for its
wider range in RM amplitudes from $+50\ \radm$ to $+300\
\radm$. Region C is in the same area as the bottom of Loop I (a circle
with diameter $116\degr \pm 4\degr$ centered around $l=329\degr \pm
1\fdg5$, $b=+17\fdg5 \pm 3\degr$ Berkhuijsen et al. 1971), but
\citet{simard1980} considered this a coincidence.
Figure~\ref{Ha_RM-fig} shows a distinct area inside region C with
consistently high positive RM in the area $33\degr < l < 65\degr$,
$0\degr < b < 40\degr$. The median RM in this area is $66.7\ \radm$,
while the median RM of the adjacent areas $13\degr < l < 33\degr$ and
$60\degr < l < 80\degr$, both with $0\degr < b < 40\degr$ is $24.6\
\radm$. The excess RM amplitude over the immediate surroundings is
therefore $42\ \radm$.  The high-latitude boundary of this region of
enhanced positive RM is not sharp. If it extends well north of
$40\degr$ at the level of $\sim 10\ \radm$, it may be responsible for
the RM excess associated with the Hercules supercluster by
\citet{xu2006}.

\begin{figure*}
\center
%\resizebox{\textwidth}{!}{\includegraphics[angle=0]{f3.eps}}
\caption{ {\bf Figure available as separate jpg file}
Correlation between RM and 0.25 keV diffuse X-ray intensity from \citet{snowden1998}. 
\label{ROSAT_RM-fig}
}  
\end{figure*}

\begin{figure*}
%\resizebox{\textwidth}{!}{\includegraphics[angle=0]{f4.eps}}
\caption{ {\bf Figure available as separate jpg file}
RM structure on the sky centered at
$(l,b)=(180\degr,0\degr)$, leaving out data with $|RM| < 25\ \radm$,
with positive RM in red and negative RM in blue, and no scaling of the
circles according to RM amplitude.  This representation is analogous
to showing a the lowest contour in a radio map at the level of a few
sigma.  It visualizes RM structure and boundaries where the sign of RM
changes. Regions with a high RM amplitude appear more densely
sampled. Note the arc-like structures in RM on angular scales of tens
of degrees that add a few hundred $\radmsq$ to the SF amplitude on
large angular scales.
\label{conditional_RM-fig}
}  
\end{figure*}

Figure~\ref{Ha_RM-fig} also shows enhanced RM amplitude associated
with some bright \HII\ regions, e.g. at $(l,b)$ = $(+105\degr,+4\degr)$,
$(+78\degr,+2\degr)$, $(+6\degr,+23\degr)$ ($\zeta$ Oph),
$(-108\degr,-9\degr)$ (Gum), and $(-165\degr,-12\degr)$ ($\lambda$
Ori). Smaller excesses of RM amplitude are associated with more
diffuse, low-surface brightness H$\alpha$ emission, e.g. in the
Orion-Eridanus region ($-180\degr < l < -135\degr$, $-55\degr < b < -5
\degr$). We also see a small RM excess associated with the shell of the 
Orion-Eridanus superbubble \citep{heiles1976,brown1995}. 

Figure~\ref{ROSAT_RM-fig} shows RMs on an image of 0.25
keV diffuse X-ray emission from \citep{snowden1997}. At low latitudes
there is little emission because of strong absorption.  At high
latitudes, emission from the more distant halo is observed, while
high-latitude clouds absorb the halo emission locally, and appear as
dark silhouettes known as X-ray shadows
\citep{snowden1997,snowden1998}. Region C$'$ coincides with a large X-ray
shadow that has an HI counterpart at $V_{\rm LSR} = -2\ \kms$, and that
is part of the outer HI shell of radio loop I \citep{berkhuijsen1971}.

\section{RM structure functions}
\label{strucfunc-sec}

Structure in RM on angular scales of tens of degrees becomes visible
because of the high sampling density ($> 1$ RM per square degree).
The RM density on the sky in our data is not sufficient to recognize
RM structures on angular scales less than a few degrees, but structure
on small angular scales does affect the variance of RM because it
introduces differences from one line of sight to another. This
addition to the variance of RM may depend on angular scale, and is
conveniently expressed as the second-order SF $D$
defined as
\begin{equation}
D(\delta \theta)= {1 \over {N} }\sum_i [ RM(\theta)-RM(\theta+\delta\theta) ]_i^2,
\label{SF-eq}
\end{equation}
where the sum is over all pairs of sources with a mutual distance
$\delta \theta$ on the sky, and $N$ is the number of pairs included in
the sum. The mean distance between an RM measurement and its closest
neighbor in our data is $0\fdg65$. SFs can probe RM variance on
smaller angular scales than this depending on the number of source
pairs available to evaluate the variance with reasonable
confidence. For our analysis, the smallest angular scale is taken to
be the scale for which we have a minimum of 20 source pairs, or
$0\fdg1$, whichever comes first. The largest angular scale we consider
for the SFs is $\delta\theta=10\degr$, so the RM gradient related to
the large-scale magnetic field does not contribute significantly to D.

\begin{figure}
\center
\resizebox{\columnwidth}{!}{\includegraphics[angle=0]{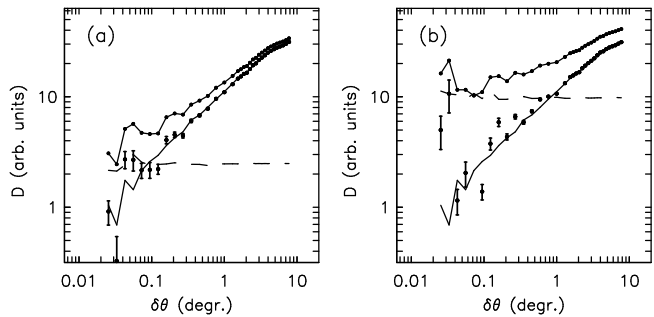}}
\caption{ Simulation of the effect of noise on a SF, assuming two
different levels of RM errors. The upper solid curves with dots
represent the measured SFs with noise. The dashed
curves show the power calculated from the adopted errors. The points
with error bars represent the SF after subtracting the noise
power. The SF made with the original 1978 noiseless data points is
represented by the lower solid curve.  The measurement errors in (b)
are on average two times as large as the measurement errors in (a),
resulting in a noise power that is 4 times higher in (b). The relative
strength of the noise power in (a) is more representative for the data
in this paper.
\label{SF_sim-fig}
}  
\end{figure}

All possible sources of variance contribute to the amplitude of the
SF. We may write D as a sum of variance terms that we consider
statistically independent
\begin{equation}
D = \sigma^2_{\rm int} + \sigma^2_{\rm IGM}(\delta\theta) + \sigma^2_{\rm ISM}(\delta\theta) + \sigma^2_{\rm noise}(\delta\theta),
\label{variance-eq}
\end{equation}
with $\sigma^2_{\rm int}$ the variance of RM originating inside the
source, $\sigma^2_{\rm IGM}$ the variance in RM originating in the
intergalactic medium, for different lines of sight separated by angle
$\delta\theta$, $\sigma^2_{\rm ISM}$ the variance of RM originating in
the interstellar medium for lines of sight separated by angle
$\delta\theta$, and $\sigma^2_{\rm noise}$ the variance resulting from
measurement errors and noise. 

The variance $\sigma^2_{\rm noise}$ should be subtracted from $D$ to
obtain the sum of the three terms of astrophysical interest
\citep[e.g.][]{haverkorn2004}. This term was calculated from the
errors on the RMs contributing to pairs separated by angular distance
$\delta\theta$. Each term in the summation of Equation~\ref{SF-eq}
contains differences $RM_m - RM_n$ ($m > n$ indices of data in the RM
catalog). For each pair of sources included in the summation, the
variance added by noise is $\sigma_m^2+\sigma_n^2$ where $\sigma_m$
and $\sigma_n$ are the standard errors for $RM_m$ and $RM_n$
respectively. This assumes that the error bars in the RM data base
provide a realistic estimate of the measurement errors. We defer this
discussion to Section~\ref{discussion-sec}.  In principle
$\sigma^2_{\rm noise}$ can depend on $\delta\theta$, but for the SFs
presented here, it is virtually constant. The median error in our
complete RM catalog is $\sim 10.8\ \radm$, so the error term can be a
significant source of variance and it affects the slope of the SF on
small angular scales.  Unless stated otherwise, SFs in this paper or
derived properties such as amplitude and slope have been corrected for
variance associated with measurement errors.

Figure~\ref{SF_sim-fig} shows a simulation of the effect of noise on a
SF, and the noise-subtracted SF.  The simulated data consisted of 1978
points drawn at random from an image with a random field with input SF
with slope 0.6 kindly provided by V. Uritsky. For
Figure~\ref{SF_sim-fig}a, measurement errors for each data point were
simulated by first drawing a standard deviation from a uniform
distribution between 0.4 and 2.0 (data units), then drawing a number
from a Gaussian distribution with the chosen standard deviation, to be
added to the actual data value.  The standard deviation itself was
recorded as the error bar for each RM. The figure shows the SF 
as observed (upper solid curve with dots), the noise power
$\sigma^2_{\rm noise}$ evaluated from the error bars of RM (dashed
curve), and the SF corrected for noise power (points
with error bars), compared with the SF of the same
1978 noiseless data points (lower solid curve).  For
Figure~\ref{SF_sim-fig}b, the same set of original data points was
used, but the standard deviations of the RM errors were drawn
independently from a uniform distribution between 0.8 to 4.0 data
units. This on average doubles the measurement errors compared with
the situation shown in Figure~\ref{SF_sim-fig}a, and quadruples the
level of the noise power in Equation~\ref{variance-eq}.

We retrieve the input SF in the simulation shown in
Figure~\ref{SF_sim-fig} and other simulations that included SFs of
various slopes and noise levels. These simulations confirmed our
evaluation of the noise power even though the errors in individual RMs
varied by a factor 5. On the smallest angular scales, we see
occasional large variation in the data after noise subtraction.  This
is to be expected, because the contribution of the noise to the total
variance is large, but our estimate of the noise power is uncertain
because relatively few source pairs with small angular separation
exist in the catalog.  We can easily identify such points and remove
them from consideration.

The subtraction of the SF of the noise is a statistical correction
that is more accurate for larger samples.  The SF of the measurement
errors is not necessarily flat, for instance in the case of periodic
variations in the noise of a mosaicking survey with widely separated
field centers, or in the case of a deep survey of a small area of the
sky where the noise increases towards the edge of the survey area
(affecting all source pairs with large angular separation). In our
analysis we do {\it not} assume that the SF of the noise is flat, or
that all RM error are equal. We calculate the SF of the noise for each
$\delta\theta$ from the actual error bars of all included
RMs. However, we found no systematic variation of the SF of the noise
with angular separation for the \citet{taylor2009} RM catalog.

\subsection{Direction dependence of the RM SF}

\begin{deluxetable*}{lrrrrrrrl}
\tablecolumns{9}
\tablewidth{0pc} 
\tabletypesize{\small}
\tablecaption{ Regions and parameters for SFs shown in Figures~\ref{SFpoles-fig},~\ref{SFregions-fig}, and ~\ref{SFGum-fig} }
\tablehead{
\colhead{Region} & \colhead{$l_{\rm min}$}  & \colhead{$l_{\rm max}$}  & \colhead{$b_{\rm min}$}   & \colhead{$b_{\rm max}$}  & \colhead{$\log(A_2)$}  & \colhead{$\alpha_1$}   &  \colhead{$\alpha_2$}  & \colhead{Figure}\\
      & \colhead{(degr.)}  & \colhead{(degr.)}  & \colhead{(degr.)}   & \colhead{(degr.)}  &        &      &     &  
}  % end tablehead
\startdata
NGP            &  $-180$  &  180    &   75   &    90  & 2.40 (0.03)  & 0.63 (0.39) & 0.05 (0.04)  & ~\ref{SFpoles-fig}a,b \\
SGP            &  $-180$  &  180    &  $-90$ &  $-75$ & 2.67 (0.03)  & 0.99 (0.58) & 0.02 (0.04)  & ~\ref{SFpoles-fig}c,d \\
Orion          &  $-180$  & $-135$  &  $-50$ &  $-10$ & 2.92 (0.02)  & 0.79 (0.08) & 0.50 (0.03) & ~\ref{SFregions-fig}a (black) \\
anti-Orion     &  $-180$  & $-135$  &   10   &   50   & 2.67 (0.01)  & 1.10 (0.39) & 0.31 (0.01) & ~\ref{SFregions-fig}a (gray) \\
Region C$'$      &    33    &   68    &   10   &   35   & 3.06 (0.01)  & 0.33 (0.20) & 0.39 (0.02) & ~\ref{SFregions-fig}b (black) \\
anti-Region C$'$ &    33    &   68    &  $-35$ &  $-10$ & 3.15 (0.03)  & 0.31 (0.48) & 0.58 (0.05) & ~\ref{SFregions-fig}b (gray) \\
Region A$'$      &    80    &  150    &  $-40$ &  $-20$ & 2.81 (0.01)  & 0.82 (0.30) & 0.51 (0.02) & ~\ref{SFregions-fig}c (black) \\
anti-Region A$'$ &    80    &  150    &   20   &   40   & 2.68 (0.01)  & 0.74 (0.26) & 0.38 (0.03) & ~\ref{SFregions-fig}c (gray) \\
Fan            &   115    &  160    &    5   &   25   & 2.78 (0.02)  & $-$0.02 (0.12) & 0.58 (0.03)  & ~\ref{SFregions-fig}d (black) \\
anti-Fan       &   115    &  160    & $-25$  &  $-5$  & 2.98 (0.02)  & 0.77 (0.17) & 0.59 (0.04)  & ~\ref{SFregions-fig}d (gray) \\
Gum            &   100    &  100    &   20   &   20   & 3.72 (0.04)  & 1.08 (0.41) & 1.14 (0.09)  & ~\ref{SFGum-fig} \\
\enddata
\label{SFregions-tab}
\end{deluxetable*}

\begin{figure*}
\center
\resizebox{\textwidth}{!}{\includegraphics[angle=0]{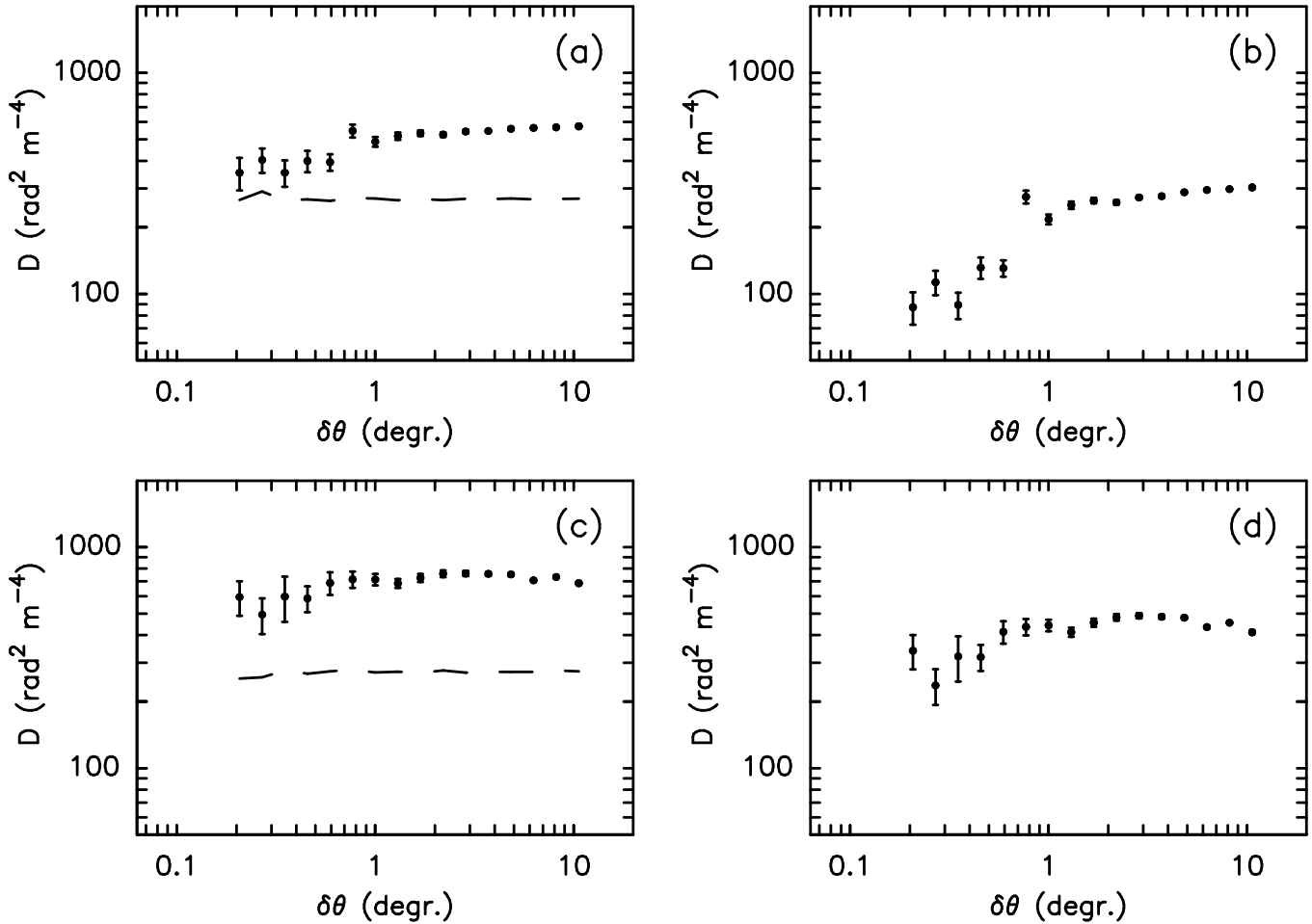}}
\caption{(a) RM SF in the NGP region ($b > 75\degr$).  The dashed line
gives the power in the SF from measurement errors. (b)
SF for the NGP region with noise power subtracted.  (c) Same as (a)
for the SGP region ($b < -75\degr$). (d) Same as (b) for the SGP
region.
\label{SFpoles-fig}
}  
\end{figure*}

\begin{figure*}
\center
\resizebox{\textwidth}{!}{\includegraphics[angle=0]{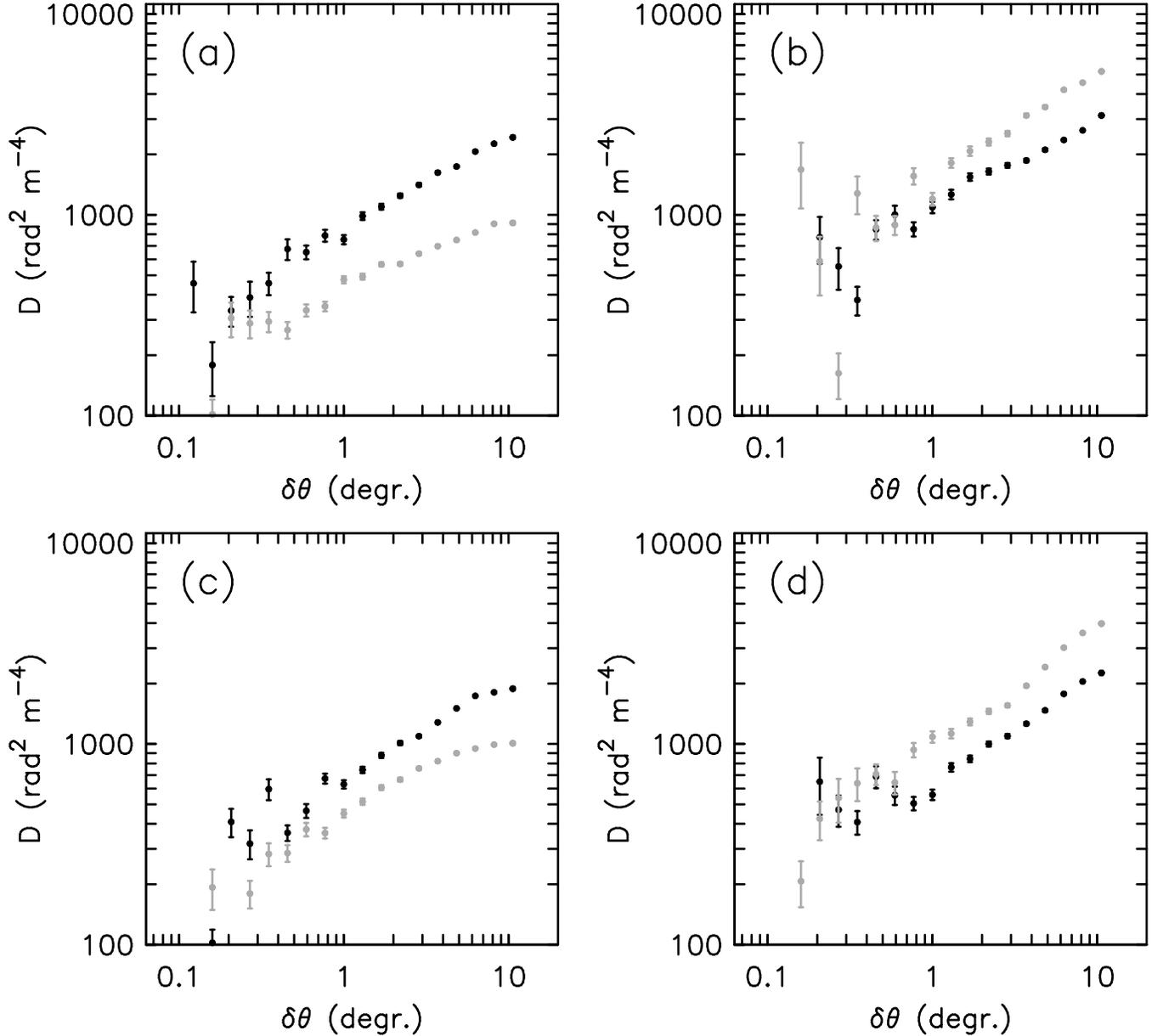}}
\caption{(a) RM SF in the Orion region (black) and its northern
counter part (``anti-Orion''; grey) as defined in
Table~\ref{SFregions-tab}. The noise variance $\sigma^2_{noise}$ has
been subtracted.  (b) SF for Region C$'$ (black) and its southern
counterpart ``anti-region C$'$'' (grey).  (c) SF for Region A$'$
(black) and its northern counterpart ``anti-region A$'$'' (grey).  (d)
SF for the Fan region (black) and its southern counterpart
``anti-Fan'' (grey).
\label{SFregions-fig}
}  
\end{figure*}

\begin{figure}
\center
\resizebox{\columnwidth}{!}{\includegraphics[angle=0]{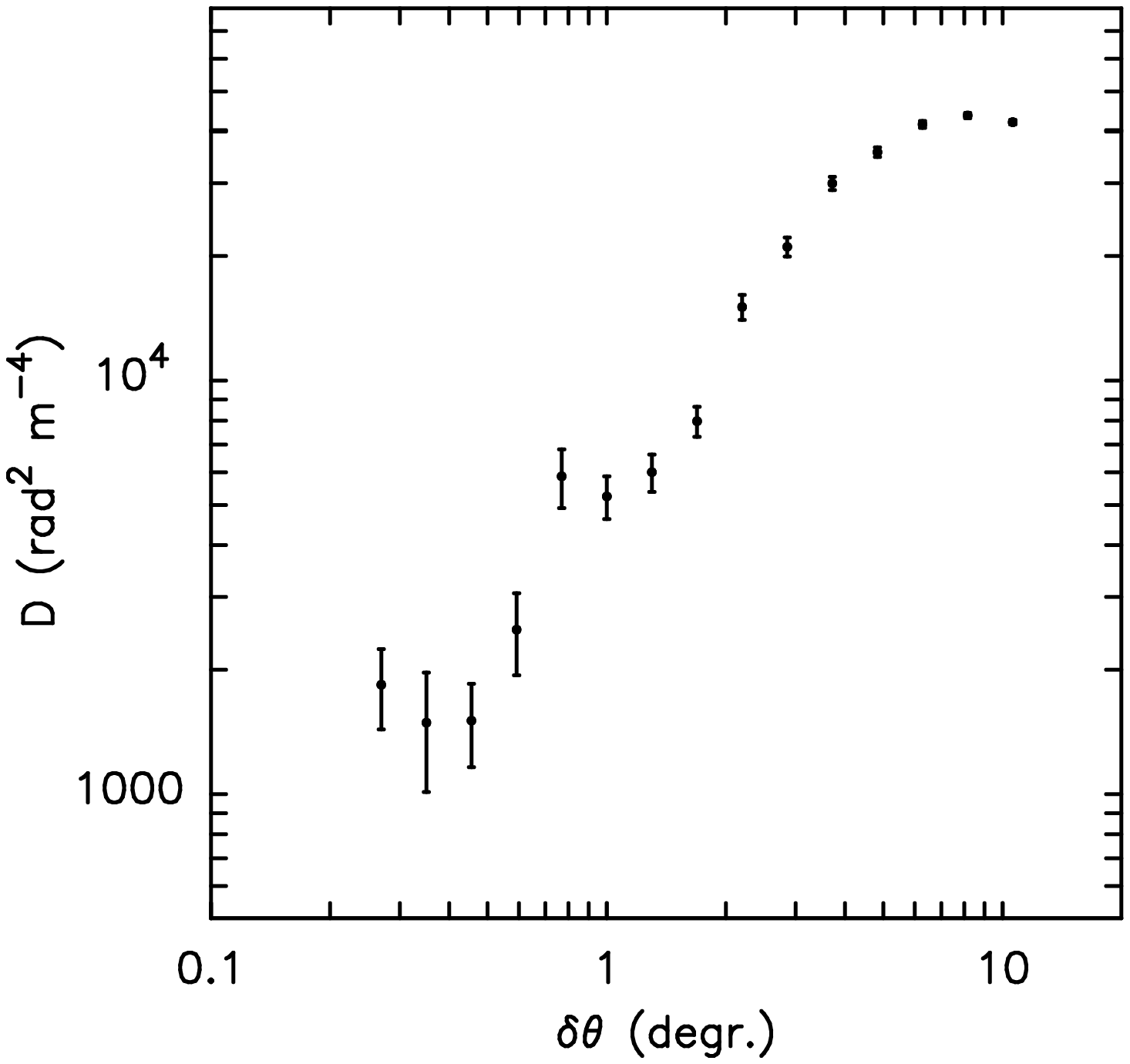}}
\caption{ SF for the region of the Gum nebula defined
in Table~\ref{SFregions-tab}.
\label{SFGum-fig}
}  
\end{figure}

Before we derive an all-sky image of SF parameters, we examine SFs for
a few areas of interest inspired by results discussed in the previous
section. These regions visualize the significance of differences in SF
parameters with direction.  Table~\ref{SFregions-tab} lists these
regions, and SFs are shown in
Figures~\ref{SFpoles-fig},~\ref{SFregions-fig}, and~\ref{SFGum-fig}.

Experience from the literature
\citep{haverkorn2007,haverkorn2008,roy2008} and from the present data
\citep{stil2009} shows that the slope of the SF is not always
constant, but sometimes changes slope at an angular scale $\delta
\theta \approx 1\degr$. The precise angular scale at which a break in
the power law occurs is not necessarily the same everywhere, but
visual inspection of SFs across the sky suggests that {\it if} a break
occurs, the SF slope is considerably higher for $\delta\theta <
1\degr$ than for $\delta\theta > 1\degr$.  We therefore fit separate
power laws to the SFs at angular scales $\delta \theta
\le 1\degr$ and $\delta \theta > 1\degr$:
\begin{equation}
D(\delta\theta) = A_i\  \delta\theta^{\alpha_i}
\end{equation}
with $i=1$ for $\delta\theta \le 1\degr$ and $i=2$ for $\delta\theta >
1\degr$. Table~\ref{SFregions-tab} lists results of such fits for the
SFs shown in Figures~\ref{SFpoles-fig},~\ref{SFregions-fig},
and~\ref{SFGum-fig}.  The fits for small angular scales are much less
constrained because of the limited number of close pairs, and a
relatively larger contribution of $\sigma^2_{\rm noise}$ to the total
variance on small angular scales. We use the fit parameter $A_2$ as a
measure of the amplitude of the SF. $A_2$ is the
variance of RM for an angular scale of $1\degr$, corrected for
variance due to measurement errors.

\subsubsection{Polar regions}

Figure~\ref{SFpoles-fig} shows RM SFs within 15 degrees of the North
Galactic Pole (NGP; 1019 sources) and the South Galactic Pole (SGP;
752 sources). Parameters for these SFs are listed in
Table~\ref{SFregions-tab}. We have fewer RMs in the SGP region because
of the declination limit of the NVSS ($\delta > -40\degr$), and
because of less sky coverage with boths IFs required to derive a RM.
The contribution of $\sigma^2_{\rm noise}$ is indicated by the dashed
curves, and subtracted in Figure~\ref{SFpoles-fig}b and d.
Subtracting the noise variance that is approximately independent of
angular scale $\delta \theta$, reduces the amplitude of the SF
significantly, and also steepens the slope of the SF on small angular
scales.

The amplitudes of the SFs at the NGP and the SGP are signifcantly
different. The amplitude of the SF in Figure~\ref{SFpoles-fig}a on
angular scales more than $2\degr$ is consistent with the amplitude
$\sim 480\ \radmsq$ found by \citet{simonetti1984} for this region.
These authors did not subtract the $\sigma^2_{\rm noise}$ term in
Equation~\ref{variance-eq}.  The SF at the NGP also shows a slight
increase in amplitude from angular scales $0\fdg2$ to $10\degr$, but
the error of $\alpha_2$ in this region is large
(Table~\ref{SFregions-tab}). \citet{mao2010} reported a slope $0.08
\pm 0.01$ in the NGP region and $0.03 \pm 0.01$ in the SGP region. Our
SF at the SGP appears flatter than at the NGP.

A difference in the slope and amplitude of the SFs of the two Galactic
pole regions suggests a significant Galactic foreground in these
regions. Assuming $\sigma^2_{\rm IGM}$ is negligible, and
$\sigma^2_{\rm int}$ is independent of $\delta\theta$ for unrelated
sources, a rise in the noise-corrected SF or differences in amplitude
must be related to differences in $\sigma^2_{\rm ISM}$. Since the
amplitude of the SF at the SGP is higher than at the NGP, and the SF
at the NGP appears to have a positive slope on smaller angular scales,
it follows that both Galactic poles have significant foreground RM
variations. An upper limit $\sigma_{\rm int} \lesssim 10\ \radm$ is
suggested by the RM variance in the NGP region on angular scales less
than $1\degr$. We do not expect $\sigma_{\rm int}$ to be much less
than $5\ \radm$, because the variation of RM across individual sources
at high latitude on angular scales less than $1\arcmin$
\citep{leahy1987} is not much smaller than our upper limit determined
at angular scales $10\arcmin$ - $25\arcmin$ from
Figure~\ref{SFpoles-fig}b, while models of the foreground suggest
lower values \citep{sun2009}. We adopt the value $\sigma_{\rm int} =
7\ \radm$ and find $\sigma^2_{\rm ISM} = 202\ \radmsq$ at the NGP, and
$\sigma^2_{\rm ISM} = 419\ \radmsq$ at the SGP. If this additional
variance arises in a single Faraday screen, the rms RM amplitude of
this screen would be $14.7\ \radm$, but there is no evidence that the
higher variance results from a separate additional Faraday screen in
the southern hemisphere. \citet{taylor2009} and \citet{mao2010}
reported that the mean RM near near the SGP is larger than near the
NGP. The difference in RM variance found here is consistent with the
difference in mean RM in the sense that both variance and mean are
larger by a factor $\sim 2$ in the south.

\subsubsection{RM SFs at lower latitude}

\begin{figure*}
\center
%\resizebox{\textwidth}{!}{\includegraphics[angle=0]{f9.eps}}
\caption{ {\bf Figure available as separate gif file}
Direction dependence of SF amplitude $A_2$ and slope
$\alpha_2$.  Top: amplitude for $\delta\theta = 1\degr$. Contour
levels are 300, 600, 1200, 2400, 4800, 9600, and 19200 $\radmsq$. The
lowest contours show tickmarks in the direction of smaller
amplitude. Grayscales represent H$\alpha$ intensity in Rayleigh from
\citet{finkbeiner2003} on a logarithmic scale indicated by the color
bar. The closed black loop marks the lower southern boundary of the
NVSS ($\delta=40\degr$), and the gray loops indicate radio loops I-IV
from \citet{berkhuijsen1973}.  Bottom: Slope of the SF on scales $> 1
\degr$ ($\alpha_2$).  Contour levels range from 0.2 to 1.2 with
increments 0.2. Contours at the levels 0.2, 0.4, and 0.8 are
labeled. The grayscales are the same as in the upper panel. The SF
slope varies between $-0.07$ and $1.28$.
\label{SF2_skydist_Ha-fig}
}  
\end{figure*}

\begin{figure*}
\center
%\resizebox{\textwidth}{!}{\includegraphics[angle=0]{f10.eps}}
\caption{ {\bf Figure available as separate gif file}
Same as Figure~\ref{SF2_skydist_Ha-fig}, with gray scales representing polarized
intensity of diffuse emission at 1.4 GHz from \citet{wolleben2006}.
\label{SF2_skydist_PI-fig}
}  
\end{figure*}

Figure~\ref{SFregions-fig} and Figure~\ref{SFGum-fig} show SFs of
selected intermediate-latitude areas defined in
Table~\ref{SFregions-tab}. The pairs of SFs in each panel of
Figure~\ref{SFregions-fig} represent regions with the same longitude,
mirrored with respect to the Galactic plane. The Gum region is shown
separately in Figure~\ref{SFGum-fig} because its mirror region is not
covered by the NVSS, and the amplitude of the SF is much larger than
the SFs shown in Figure~\ref{SFregions-fig}. The first impression of
Figure~\ref{SFregions-fig} is that the amplitude of the SFs is not
symmetric with respect to the Galactic plane. The highest SF in each
panel of Figure~\ref{SFregions-fig} is in the southern Galactic
hemisphere, showing the same asymmetry as Figure~\ref{SFpoles-fig}.
On average these SFs suggest that the variance of RM in different
areas in the southern Galactic hemisphere is $\sim 50\%$ larger than
in the northern Galactic hemisphere. \citet{taylor2009} found that the
mean RM across the southern Galactic hemisphere is consistently higher
than in the northern Galactic hemisphere.  We will show that the
asymmetry in RM variance is also global.

Figure~\ref{SFregions-fig} and Figure~\ref{SFGum-fig} also show
significant differences in the slope of the SFs. The SF of the Gum
region is among the steepest identified in the data. The Orion
region has a significantly steeper slope than its counterpart region
across the Galactic plane, and the same is true for Region A$'$. In
other areas, such as the Fan region, we see a difference in amplitude
but no significant difference in slope.

Figure~\ref{SF2_skydist_Ha-fig} the sky distribution of $A_2$ and
$\alpha_2$ derived from a dense grid of SFs, using sources inside a
circular aperture with diameter $15\degr$. The slope and amplitude of
the SFs for $\delta\theta > 1\degr$ have a similar accuracy as those
listed in Table~\ref{SFregions-tab}, because of the similar number of
sources involved.  The uncertainty in the amplitude is therefore
approximately 5\%, and the uncertainty in the slope is approximately
0.03. Most of the structure visible in Figure~\ref{SF2_skydist_Ha-fig}
represents significant differences in amplitude or slope of the RM SF.

The asymmetry in SF amplitude between the NGP and SGP shown in
Figure~\ref{SFpoles-fig} is also visible in the top panel of
Figure~\ref{SF2_skydist_Ha-fig}. There are only a few areas in the
south where the amplitude drops below $300\ \radmsq$, while most of
the NGP region has amplitudes less than $300\ \radmsq$. Towards the
Galactic plane, RM fluctuations increase. The largest fluctuations
($4.8 \times 10^5\ \radmsq$) are found in the Galactic plane, in the
direction $l \approx 90\degr$. RM fluctuations may be approximately
symmetric with respect to $l = 0\degr$, but the declination limit of
the NVSS removes the fourth Galactic quadrant and part of the third
Galactic quadrant.

The amplitude of the SFs in Figure~\ref{SF2_skydist_Ha-fig}a traces
the brightest extended H$\alpha$ emission, with significantly higher
amplitudes towards the inner Galaxy than towards the outer
Galaxy. This contrast shows the excentric location of the Sun in the
Galaxy, and thus suggests that the low-latitude RM variance on an
angular scale of $1\degr$ originates from a line of sight distance of
several kpc. Structure at higher latitudes are probably more local in
nature.  Around longitude $l = -110\degr$, the contours of SF
amplitude follow the outline of the Gum nebula in the H$\alpha$
image. The low-latitude peaks follow bright emission in the Galactic
plane, except perhaps the local maximum near $l = 40\degr$. The
brighter parts of the Orion-Eridanus region ($l=-170\degr$,
$b=-40\degr$) are also associated with enhanced RM variance. The
outlines of radio Loops I-IV \citep{berkhuijsen1973} are also shown in
Figure~\ref{SF2_skydist_Ha-fig}.  Only Loop III may have have a
counterpart in SF amplitude.

The slope of the SF (Figure~\ref{SF2_skydist_Ha-fig}b) is highest in
the Galactic plane, but it does not show the same concentration toward
the Galactic plane as the amplitude. The largest slopes are $\alpha_2
\approx 1.3$ in the direction of the $\zeta$Oph \HII\ region
($l=5\degr$, $b=+24\degr$), and the Gum nebula. At higher northern
Galactic latitude, we see a systematic trend that the slope of the SF
is shallower towards the Galactic center than towards the Galactic
anticenter at the same latitude. The contour of $\alpha_2 = 0.2$
reaches $b\approx 50\degr$ towards $l=0\degr$, while it is found
around $b \approx 80\degr$ towards $l = 180\degr$.  It is not clear
whether the same behavior also occurs at southern Galactic latitudes,
because of the unsampled region $\delta<-40\degr$. The $\alpha_2=0.2$
contour is found mostly between $b=-60\degr$ and $b=-70\degr$, except
for $0\degr < l < 30\degr$ where it is closer to the Galactic plane
similar to its northern counterpart.  The over-all distribution of the
SF slope at high latitudes is more symmetric with respect to the
Galactic plane than the amplitude. There is a suggestion of a steeper
slope in the rim of radio Loop III.
 
Figure~\ref{SF2_skydist_PI-fig} shows the amplitude and slope of RM
SFs as contours on polarized intensity of diffuse
emission at 1.4 GHz from \citet{wolleben2006}. The main features in
this polarized intensity were described by \citet{wolleben2007}. The
gradual trend in slope if high-latitude SFs with longitude mimics the
boundaries of the high-latitude polarized emission described by
\citet{wolleben2007}.

In the first Galactic quadrant ($0\degr < l < 90\degr$), the drop in
polarized intensity at $b \approx 30\degr$ and $b \approx -30\degr$
noted by \citet{wolleben2006} corresponds with RM variance of $600 \pm
100\ \radmsq$ north and south of the Galactic plane, although we see
extended polarized emission in some areas with RM variance up to $900\ 
\radmsq$. The polarized intensity drops abruptly by a factor $\sim 3$
at the boundary, while the average percentage polarization is a factor
$\sim 10$ smaller at lower latitudes than for $b > 30 \degr$
\citep{wolleben2007}. Some of this difference may result from
depolarization at a larger distance. The 23 GHz WMAP polarization
image \citep{page2007} shows the North Polar Spur polarized at lower
latitudes, implying some depolarization occurs for $|b| < 30\degr$ at
1.4 GHz. We take the abrupt decrease of a factor $\sim 3$ in polarized
intensity as a measure for the amount of depolarization by a possible
foreground screen.

In the second Galactic quadrant ($90\degr < l < 180\degr$), the
brightest polarized emission is observed at lower latitudes north of
the Galactic equator in the Fan region. The brightest polarized
emission of the Fan region is intersected by the $900\ \radmsq$ RM
variance contour. In contrast to the first quadrant we do not see a
clear anti-correlation between polarized intensity and RM variance,
with the possible exception of Radio Loop III.

The third Galactic quadrant ($180\degr < l < 270\degr$) is only
partially sampled by the present data. The region of the Gum nebula as
traced by the $600\ \radmsq$ contour appears depolarized. A narrow
polarized feature between longitudes $-180\degr > l > -155\degr$, and
latitude $-10\degr < b < 0\degr$ is an extension of polarized emission
in the second quadrant. It does not appear to be correlated with
structure in RM variance.

\section{Discussion}
\label{discussion-sec}

\subsection{RM contributions from the local ISM}

\begin{figure*}
\center
\resizebox{14cm}{!}{\includegraphics[angle=0]{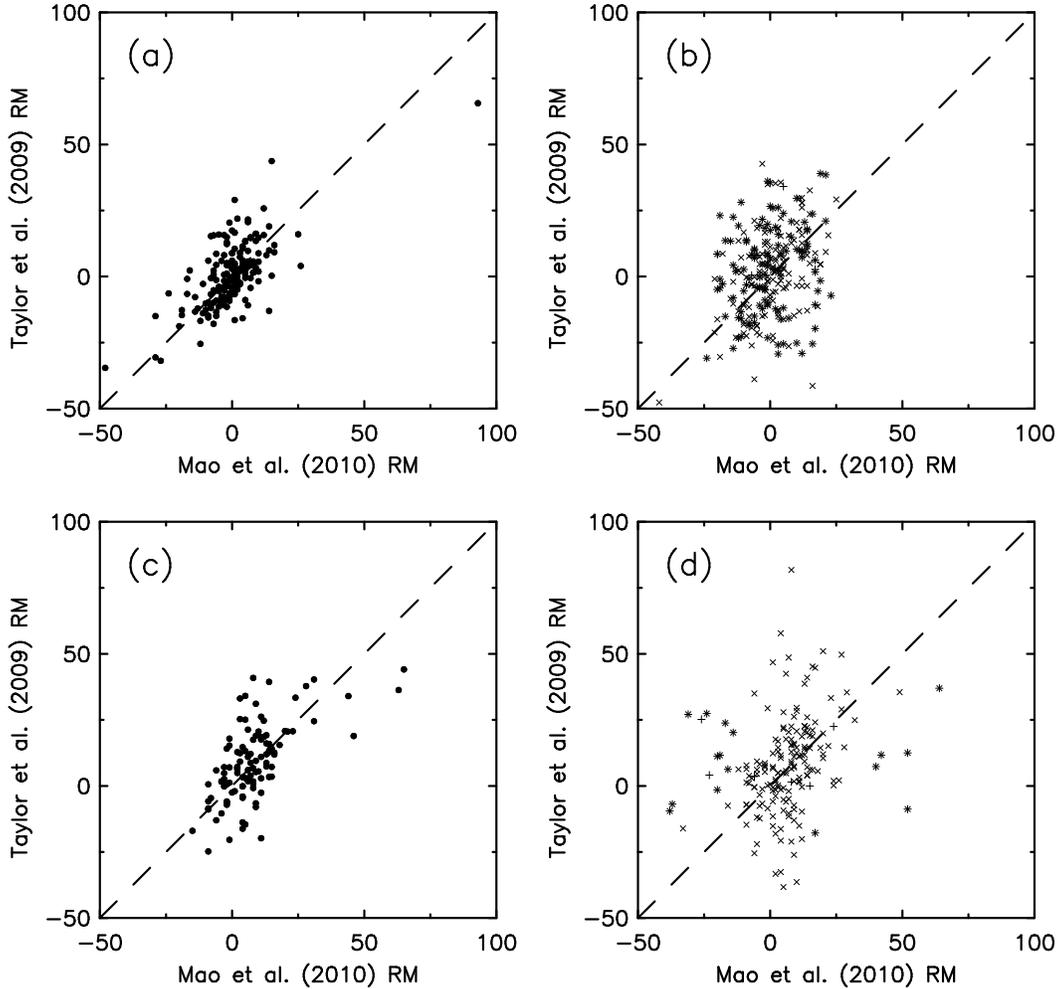}}
\caption{ Comparison of RMs from \citet{taylor2009} with RMs from
\citet{mao2010}, divided by stated RM error. (a): NGP region with RM
error $< 8\ \radm$ in both samples. (b): NGP region with RM error $\ge
8\ \radm$ in \citet{taylor2009} ($\times$), and RM error $\ge 8\
\radm$ in \citet{mao2010} ($+$).  If the errors in both surveys are
$\ge 8\ \radm$ both symbols appear, resulting in a $*$. (c): same as
(a) for the SGP region. (d): same as (b) for the SGP region.
\label{RM_RM_comp-fig}
}  
\end{figure*}

RMs from polarized extragalactic sources probe the entire line of
sight from the source to the observer. Most extragalactic sources
experience relatively little internal Faraday rotation, and
intergalactic Faraday rotation is believed to be substantially smaller
than the intrinsic RM variance of the extragalactic sources.  With the
exception of some sources with high internal Faraday rotation, most of
the RM originates from the section of the line of sight inside our own
Galaxy. RM amplitude also increases significantly towards the Galactic
plane. Although the RM is an integral along the line of sight, a
single object can contribute significantly.  The Gum nebula
\citep[distance $\sim 400$ pc][]{brandt1971} contributes significantly
to the RM, even though the line of sight towards the Gum nebula
intersects the inner Galaxy, where the electron density is high over a
line of sight of several kpc \citep{cordes2002}. The Orion star
forming region, also at a distance of approximately 400 pc
\citep{dezeeuw1999}, is at higher latitude, so its background is a
relatively short line of sight through the Galaxy.

The line of sight through the Galaxy is much shorter for high-latitude
sources, and structures in the local ISM may affect the RM
significantly. Some high-latitude \HII\ regions were identified as
such, consistent with earlier results that showed that individual
\HII\ regions can have a significant impact on RM
\citep{mitra2003}. We have found evidence that Region C$'$ is
associated with a nearby cloud revealed by its absorption of soft
X-rays from the Galactic halo that is at least partially neutral
because it is visible in \HI\ emission. This associates region C$'$
with the outer HI shell of Loop I. The enhanced RM amplitude in region
C$'$ spans only $35\degr$ in Galactic longitude, with its central
longitude around $l = 45\degr$ \citep[see also][]{stil2007}.

Towards the Galactic anticenter, where we expect almost no
contribution from the large-scale Galactic magnetic field, we find
curved filamentary structures in RM with amplitudes of a few tens
$\radm$, positive and negative (Figure~\ref{conditional_RM-fig}). 
From Equation~\ref{FR-eq}, we obtain an order-of-magnitude estimate for the
RM imposed by a structure along the line of sight with electron
density $n_{e}$, internal regular magnetic field $B$, and
line-of-sight extent $l$:
\begin{equation}
RM \approx 8\ \Bigl({n_e \over {0.1\ \rm cm^{-3}}}\Bigr) \Bigl( {B \over {1\ \mu{\rm G}}} \Bigr)\Bigl( {l \over {100\ {\rm pc}}} \Bigr)\  \radm.  
\label{RM_est-eq}
\end{equation}
The RM amplitudes and morphology of these structures resemble one or a
superposition of several bubble walls with size of the order of
$\sim$100 pc (with a large uncertainty). One of these is the
Orion-Eridanus superbubble ($170\degr < l < 220\degr$, $-30\degr < b <
-50\degr$ \citep{brown1995}, visible in emission in
Figure~\ref{Ha_RM-fig}, and in absorption in
Figure~\ref{ROSAT_RM-fig}. These filaments contribute to the amplitude
of RM SFs on large angular scales, increasing the SF slope $\alpha_2$.

The overarching picture that emerges is one of RM structure in the
local ($\lesssim$ few kpc) ISM on large scales is mostly associated
with bubble walls, \HII\ regions, and local clouds, region $A'$
(Table~\ref{SFregions-tab}) being a possible exception.  If this is
representative for an average few kpc line of sight through the
Galaxy, the RM variance added by structures on scales of a few hundred
pc should contribute significantly to the total variance in RM at low
Galactic latitudes. If bubble walls contribute significantly, models
of RM variance should include a correlation between field strength and
density on scales of a few hundred pc.

\subsection{Amplitude and slope of SFs}

\subsubsection{Comparison with Mao et al. (2010)}

\citet{mao2010} derived SFs at the NGP and SGP caps (latitude $|b| >
77\ \degr$) based on 354 and 319 RMs respectively.  \citet{mao2010}
find almost identical SF amplitudes at both Galactic poles, an
apparent contradiction of the result in Figure~\ref{SFpoles-fig}.
These authors also presented a comparison of their RMs with our RM
catalog, and foud a small correlation coefficient between the two
samples (0.39 towards the NGP, and 0.36 towards the
SGP). \citet{mao2010} suggested that the differences between the
samples are related to multiple RM components in some sources that
cannot be resolved in the two-frequency data of \citet{taylor2009},
but they did not provide specific examples where this would be the
case.  While it is true that the two-frequency data in
\citet{taylor2009} cannot resolve non-linearity in the relation
between polarization angle and $\lambda^2$, rotation measure synthesis
with limited $\lambda^2$ coverage, as in the sample of \citet{mao2010}
has its limitations \citep{brentjens2005,frick2010}.  It was recently
pointed out by L. Rudnick et al. (unpublished) that RM synthesis can
provide erroneous solutions in a simple system with just two discrete
RM components.

In order to gain a better understanding of the difference between
SFs derived by \citet{mao2010} and those derived here,
we must first compare the RM samples in more detail. The median error
for sources common to both RM catalogs is $8\ \radm$ in
\citet{taylor2009}, $5\ \radm$ in the NGP region of \citet{mao2010},
and $3\ \radm$ in the SGP region of
\citet{mao2010}. Figure~\ref{RM_RM_comp-fig} shows scatter plots of
RMs matched between the two samples, subdivided into RMs with error $<
8\ \radm$ (panels a and c), and all other RMs (panels b and d).

When considering only RMs with small errors, we see a correlation
between the two datasets at both Galactic poles
(Figure~\ref{RM_RM_comp-fig} a and c).  The larger scatter in the SGP
area found by \citet{mao2010} is related to RMs with larger error
bars, in either \citet{taylor2009}, or \citet{mao2010}, or both.  We
do {\it not} calculate a correlation coefficient, because the
correlation coefficient between independent RM surveys of low-RM
sources {\it must} be small. Unresolved multiple RM components may
affect both surveys in a similar way, and thus introduce correlated
systematic errors without increasing the scatter in
Figure~\ref{RM_RM_comp-fig}, but still raise the amplitude of
SFs derived from both data sets.

Figure~\ref{RM_RM_comp-fig} a and c both show significant RM
signal. The scatter in these relations provides an estimate of the
variance introduced by measurement errors. We examine the subsets with
small errors shown in Figure~\ref{RM_RM_comp-fig} a and c, as well as
the complete sample.  For the subsets with small errors, the total
variance is $90\ \radmsq$ in the north and $135\ \radmsq$ in the
south. The error bars (added in quadrature for each source) account
for a total variance of $37\ \radmsq$ in Figure~\ref{RM_RM_comp-fig}a
and $36\ \radmsq$ in Figure~\ref{RM_RM_comp-fig}b. For the complete
sample, the total variance is $151\ \radmsq$ in the north and $193\
\radmsq$ in the south, while the error bars account for $103\ \radmsq$
and $90\ \radmsq$ respectively.

The stated errors underestimate the true variance in the correlation
of RMs common to both samples, but more so if the stated errors are
small. The RM errors in \citet{taylor2009} were estimated by error
propagation of the noise in the Stokes $Q$ and $U$ images, while the
errors in \citet{mao2010} were estimated as half the width of the RM
spread function divided by the signal to noise ratio in
polarization. The error estimates in both surveys converge to zero as
the signal to noise ratio in polarization increases without
bound. Perhaps the true errors do not converge as rapidly as
assumed. Before we consider the effect of this underestimation of
errors on the amplitude of SF at the Galactic poles, we first consider
another aspect of the error statistics of the data.

\begin{figure}
\center
\resizebox{\columnwidth}{!}{\includegraphics[angle=0]{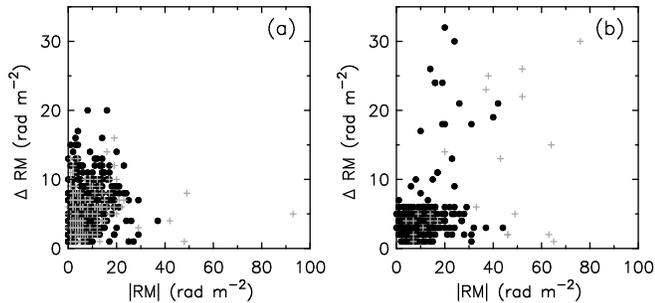}}
\caption{ Correlation of RM error with RM amplitude for the sample of
\citet{mao2010}. (a): NGP region observed with the WSRT, (b): SGP
observed with the ATCA. Black dots represent RMs used in the SF by
\citet{mao2010} (heavily saturated by overlapping symbols in some
parts of the figure). The gray crosses represent data rejected by
\citet{mao2010} for use in SFs.
\label{RM_error-fig}
}  
\end{figure}

Figure~\ref{RM_error-fig} shows the stated errors as a function of RM
amplitude for the NGP and SGP regions of the \citet{mao2010} sample.
Black dots represent RMs used in the construction of SFs, and gray
crosses respresent rejected RMs. While the median error in
Figure~\ref{RM_error-fig}a is larger, Figure~\ref{RM_error-fig}b
contains data with large RM errors that contribute to the increased
scatter in Figure~\ref{RM_RM_comp-fig}d.  RMs with amplitude more than
$20\ \radm$ in Figure~\ref{RM_error-fig}b are much more likely to have
a large error, introducing a correlation between RM error and RM
amplitude: the SGP data of \citet{mao2010} are heteroskedastic.  The
highest RM amplitudes (a few percent of the total sample) are rejected
in both regions, but the distribution of rejected RMs is more skewed
toward high RM amplitude in the south (Figure~\ref{RM_error-fig}b).

\begin{figure}
\center
\resizebox{\columnwidth}{!}{\includegraphics[angle=0]{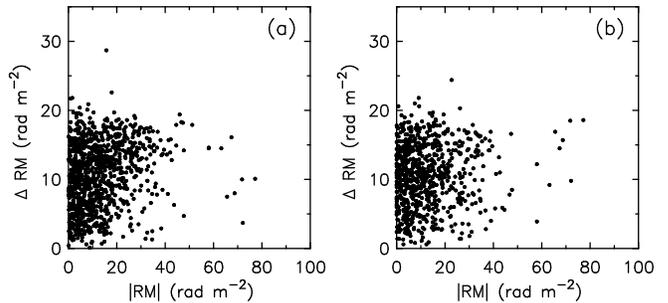}}
\caption{ Same as Figure~\ref{RM_error-fig} for the data from
\citet{taylor2009}.  (a) NGP region. (b) SGP region.
\label{Taylor_error-fig}
}  
\end{figure}

Some RMs in \citet{mao2010} were rejected in order to eliminate AGNs
with strong internal Faraday rotation, and to minimize the effect of
known Galactic and extragalactic structures. This approach is
preferred if one wishes to determine the intrinsic RM variance of
AGN-powered radio sources, or investigate small-scale turbulence at
high Galactic latitude.  This paper deals with the complete RM
foregound. Figure~\ref{Taylor_error-fig} shows the RM errors in
relation to RM amplitude for our data in the NGP and SGP regions.  The
larger RM errors in our data are compensated by the $\sim 3$ times
larger sample size, especially when probing fluctuations on small
angular scales.

What does this all mean for the SF amplitudes at the Galactic poles?
Figure~\ref{SF_poles_comp-fig} shows RM SFs at the NGP and SGP. The
gray SFs were made with the censored data of \citet{mao2010} (354 RMs
at the NGP and 319 RMs at the SGP). We find a smaller amplitude than
\citet{mao2010}, because these authors underestimated the noise power
by approximately a factor 2. The problem was identified and the
authors confirmed they recover the SF shown in
Figure~\ref{SF_poles_comp-fig} (A. Mao, private communication).  The
black points in Figure~\ref{SF_poles_comp-fig} show SFs made with the
complete sample of \citet{mao2010}. The origin of the large scatter
for $\delta\theta \lesssim 1\degr$ in Figure~\ref{SF_poles_comp-fig}a
was discussed in Section~\ref{strucfunc-sec}
(Figure~\ref{SF_sim-fig}).  Using the complete sample of
\citet{mao2010}, we confirm that the SF in the SGP region is a factor
$\sim 2$ higher than in the NGP region, but both SFs in
Figure~\ref{SF_poles_comp-fig} have a somewhat smaller amplitude than
those in Figure~\ref{SFpoles-fig}.  Both data sets suggest a small
upward slope of the SF in the NGP region.  In
Table~\ref{SFregions-tab} we list $\alpha_2 = 0.05 \pm 0.04$, while
\citet{mao2010} found $0.08 \pm 0.01$. For the SGP region we find
$\alpha_2 = 0.02 \pm 0.04$, while \citet{mao2010} found $0.03 \pm\
0.01$.

If we {\it assume} that the noise power in our SFs is really a factor
1.5 higher than that calculated from the RM errors, we would retrieve
the black curves in Figure~\ref{SF_poles_comp-fig}. This would be
equivalent with the RM errors being underestimated by a factor 1.22 in
\citet{taylor2009}, and account for the variance in the comparison of
RMs from the two surveys. While this appears a plausible explanation
of the difference, the actual comparison may be more complicated. This
analysis assumes that Figure~\ref{RM_RM_comp-fig} reveals all RM
errors, while it does not reveal possible systematic errors common to
both surveys that may increase SF amplitude.  Another question is the
nature of the sources in \citet{mao2010} for which no reliable RM
could be found (6\% of their data in the NGP region and 28\% in the
SGP region.

It is encouraging that we find consistency between SFs from
independent data sets, after accounting for all noise power traced by
direct comparison of RMs common to both surveys. This consistency
requires just a $22\%$ increase in the errors quoted by
\citet{taylor2009} (in a low-RM environment), corresponding with only
a few $\radm$.  It suggests that either the effect of multiple RM
components on the data of \citet{taylor2009} is quite small, or the
data from \citet{mao2010} are affected in the same way for most
sources. However, such a systematic error should still increase the
amplitude of the SFs, assuming it is uncorrelated between
sources. From the total amplitude of the SFs at high
latitude, we estimate that any such systematic error should be smaller
than $\sim 10\ \radm$.

Our analysis shows that both RM surveys point indicate a larger RM
variance in the direction of the SGP than in the direction of the
NGP. Our SFs at intermediate Galactic latitude show
the same trend, and these are less sensitive to the amplitude of the
noise power because it contributes much less to the total variance.
Both \citet{taylor2009} and \citet{mao2010} reported a higher mean RM
in the south than in the north. \citet{taylor2009} found this to be
the case in the entire southern Galactic hemisphere.

\subsubsection{Distribution of RM variance over the sky}

Variance in RM originates from fluctuations in electron density and
magnetic field on a range of scales in the ISM, including \HII\
regions, bubbles, turbulence, and possibly also molecular clouds with
a small ionized fraction but a strong magnetic field.  The variance in
RM should be approximately proportional to the length of the line of
sight through the Galaxy \citep{sokoloff1998}.

Figure~\ref{SF2_skydist_Ha-fig} shows that the amplitude of SFs on an
angular scale $\delta\theta = 1\degr$ follows the distribution of
H$\alpha$ intensity in the Galactic plane. The highest amplitudes at
low Galactic latitude are found in the direction of the Local Arm ($l
\approx 90\degr$) and other regions along the Galactic plane.  The
peak in RM variance at $(l,b) \approx (40\degr,0\degr$) occurs in a
region of the Galactic plane with strong extinction at visible
wavelengths. The enhance variance in the Gum region is clearly
extented in Figure~\ref{SF2_skydist_Ha-fig} with morphology similar to
H$\alpha$ intensity. The low-latitude RM variance is smaller in the
direction of the Galactic anti-center, where the line of sight through
the Galaxy is shorter. These observations are consistent with the
expectation that variance in RM increases with the length of the line
of sight through the Galaxy \citep{sokoloff1998}. Higher electron
density and stronger turbulence related to more intense star formation
in the inner Galaxy probably also contribute to the high RM variance
for lines of sight through the inner Galaxy. The outer scale in RM
variance associated with interarm regions, but not with spiral arms
reported by \citet{haverkorn2007} and \citet{haverkorn2008} indicates
that besides line-of-sight distance, the local conditions are also
important.

\begin{figure}
\center
\resizebox{\columnwidth}{!}{\includegraphics[angle=0]{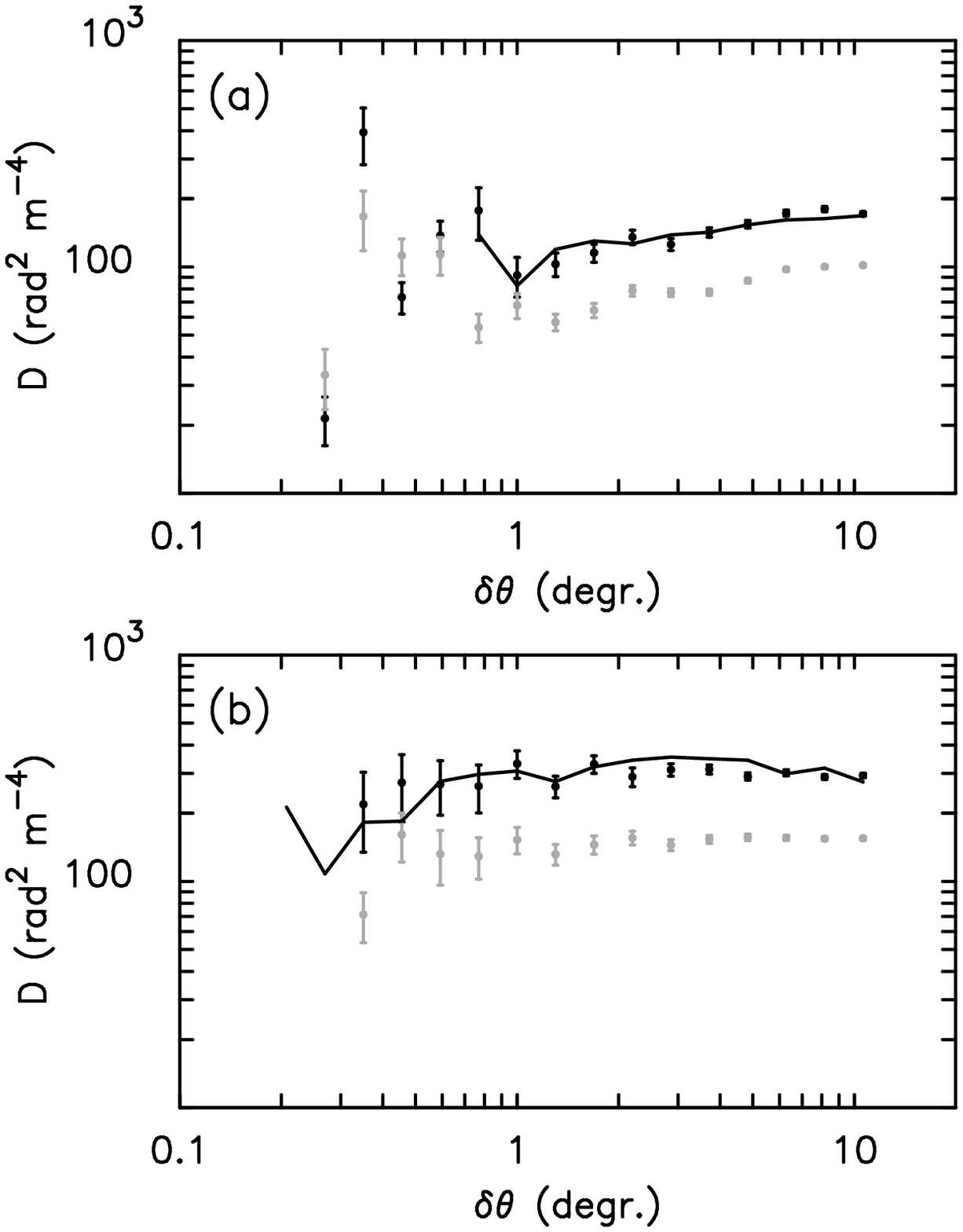}}
\caption{ SFs at the NGP (a) and SGP (b) regions, comparing results
obtained with data from \citet{taylor2009} and \citet{mao2010}. Gray
points are SFs made with RMs from \citet{mao2010}, using 354 accepted
RMs in the north and 319 accepted RMs in the south. These SFs differ
from Figure 6 in \citet{mao2010} as discussed in the text.  The black
dots represent SFs that include all RMs in \citet{mao2010}.  The black
curves represent the SFs from this paper (Figure~\ref{SFpoles-fig})
with the noise power multiplied by 1.5 before subtraction. The black
curve in panel (a) does not extend to angular scales smaller than
$0\fdg7$ because the noise power times 1.5 is comparable to the total
variance of the data.
\label{SF_poles_comp-fig}
}  
\end{figure}

The difference in RM variance between the Galactic northern and
southern hemispheres (Figure~\ref{SF2_skydist_Ha-fig} and
Table~\ref{SFregions-tab}) is particularly interesting.  Assuming the
intrinsic RM variance of extragalactic sources to be $\sigma_{\rm int}
= 7\ \radm$ (Equation~\ref{variance-eq}), the RM variance introduced
by the Galactic foreground is 202 $\radmsq$ at the NGP, and 419
$\radmsq$ at the SGP (120 $\radmsq$ and 307 $\radmsq$ respectively
using SFs from Figure~\ref{SF_poles_comp-fig}). The RM variance near
the SGP originating from the Galactic foreground is therefore a factor
2 larger than the Galactic RM variance near the NGP. The difference in
variance is equivalent with an additional source of RM variation with
amplitude $14\ \radm$ in the southern Galactic hemisphere.  The Sun is
located 20 pc from the Galactic mid-plane in the direction of the NGP.
This distance is too small compared with the scale height of the
ionized gas to make a significant difference in RM variance.  It is
more likely that we see a local asymmetry of density or magnetic
field, or both with respect to the Galactic plane. \citet{taylor2009}
also found a factor $\sim 2$ difference in the mean RM of the NGP and
the SGP, while \citet{mao2010} found a significant positive RM at the
SGP but not at the NGP.

The slope of the SF is higher near the Galactic plane with most values
in the range from 0.4 to 0.8 and excesses up to 1.3 in the direction
of two nearby \HII\ regions ($\zeta$ Oph and the Gum nebula). The
slope of our SFs (Table~\ref{SFregions-tab}) is
generally smaller than the 5/3 slope expected for Kolmogorov
turbulence \citep{armstrong1995,minter1996}, to the smallest angular
scales probed by our data ($\sim 0\fdg3$).  \citet{beck2007} found RM
SF slopes $\sim 0.3$ in the nearly face-on galaxy NGC
6946 on scales up to 6 kpc, which is in the range of values of
$\alpha_2$ in Table~\ref{SFregions-tab}.

If a change in slope occurs in the range of angular scales probed by
the current data ($0\fdg1 < \delta\theta < 10\degr$), the slope on
larger angular scales is usually smaller. A change in SF slope is
sometimes interpreted as a physical scale in the turbulence probed by
RM variance, that could be an outer scale length if the SF becomes
flat on large scales. An implicit assumption is that the RM variance
traces stochastic variations of electron density and magnetic field
related to turbulence in the magneto-ionized ISM. If deterministic
structure, e.g. an \HII\ region that may or may not be detectable in
current H$\alpha$ surveys, or a meso-scale structure in the magnetic
field, contributes significantly to the variation of RM with position
over the area for which the SF is evaluated, the shape of the SF may
reflect the angular scale of the ionized structure, not the turbulence
along the line of sight.  The present data suggest that this may be
true in some directions.  The steepest SFs at low latitude are
associated with bright extended \HII\ regions in
Figure~\ref{SF2_skydist_Ha-fig}. Figure~\ref{conditional_RM-fig} also
shows RM structure on angular scales $\gtrsim 10\degr$ with amplitude
of a few tens of $\radm$, sufficient to affect the slope of the SF
between $\delta\theta=1\degr$ and $\delta\theta=10\degr$. Sampling
with a denser RM grid can reveal such structures, as in the case of
the RM anomaly in Cygnus \citep{whiting2009}. Future wide-area RM
surveys with the Arecibo radio telescope (GALGACTS), the Australian
SKA Pathfinder ASKAP (POSSUM), and with the Square Kilometre Array
will provide denser sampling that may reveal underlying structure that
affects the slope of our SFs, while not recognizable
in the present data.

Model SFs by \citet{sun2008} and \citet{sun2009} that assume a
turbulent Kolmogorov power spectrum for density and magnetic field,
predict a flat SF at low latitude on angular scales $\delta\theta
\gtrsim 3\arcmin$ as the number of turbulent cells in the line of
sight increases. These models explore angular scales much smaller than
those probed here.  We see a general increase in SF slope towards the
Galactic plane, contrary to the expectation of a smaller slope
expected by \citet{sun2009}. As noted in
Figure~\ref{SF2_skydist_Ha-fig}, the lack of a strong longitude
dependence of SF slope, and its wider distribution with Galactic
latitude suggest that the slope of SFs on larger angular scales is
affected by the local interstellar medium.  This suggests that local
structure not represented in the models of \citet{sun2009} may be
responsible for the steep slope of the SFs at low latitude.

\subsection{Depolarization of diffuse emission}

RMs s of extragalactic sources are affected by Faraday rotation along
the entire line of sight from the source to the observer, but mostly
in the Galaxy. Depolarization of diffuse emission extends to $b
\approx 30\degr$. At this latitude, the line of sight through the
Galactic ISM is approximately twice as long as the line of sight
towards the Galactic poles. Adopting a scale height of the warm
ionized medium $H = 1.83^{+0.12}_{-0.25}$ kpc \citep{gaensler2008},
most of the Faraday rotation in a line of sight at $b = 30\degr$
should arise within a distance $\sim 3.6$ kpc. This is still more than
an order of magnitude larger than the distance to the North Polar
Spur, for which we adopt the distance to the Scorpio-Centaurus OB
association \citep[170pc;][]{egger1995}, although the near side may be
closer.  Assuming an exponential density profile, and $\delta n_e \sim
n_e$, $\delta B \sim B$ independent of density, approximately half of
the Faraday rotation and RM variance occurs in front of the North
Polar Spur. Large-scale inhomogeneity of the ISM introduces
significant uncertainty in this estimate.

A foreground screen with RM variance $\sigma_{\rm RM}^2$ on scales
much smaller than the $0\fdg$ beam size of the DRAO 26-m telescope
depolarizes emission observed at wavelength $\lambda$ by a factor
\begin{equation}
{p \over p_0} = \exp(-2 \sigma^2_{\rm RM}\lambda^4),
\end{equation}
\citep{sokoloff1998}. The sudden drop of a factor $\sim 3$ in
polarized intensity at $|b| = 30\degr$ observed at $\lambda = 21\ \rm
cm$ requires $\sigma_{\rm RM}^2 = 280 \pm 180\ \radmsq$, where the
error represents a factor 2 uncertainty in the depolarization
factor. The observed total RM variance is $600\ \pm\ 100\
\radmsq$. Around the depolarization edge, the SF slope is between 0.3
and 0.4 (Figure~\ref{SF2_skydist_PI-fig}). Assuming a typical slope
0.35 around $|b| = 30\degr$, we find the observed RM variance on the
scale of the beam of the DRAO 26-m telescope $(470\ \pm\ 100)\
\radmsq$. Subtracting the variance related to measurement errors
$\sigma_{\rm noise}^2 = 270\ \pm 18\ \radmsq$ and the intrinsic
variance of background sources $\sigma^2_{\rm int} = 49\ \pm 20\
\radmsq$ (Equation~\ref{variance-eq}), the RM variance attributed to
the Galactic ISM along the line of sight is $150\ \pm\ 110\
\radmsq$. Approximately half of this variance is expected to arise
behind the North Polar Spur, and would not depolarize emission from
this region.

The observed variance is a factor $\sim 3$ lower than what is required
to depolarize the diffuse emission, but in view of the estimated
errors, the difference is not very significant.  Within the errors, it is
possibile that differential Faraday rotation along the line of sight
contributes to depolarization of the diffuse emission.

\section{Summary and Conclusions}

This paper presents an analysis of structure in the magneto-ionized
interstellar medium probed by 37,543 RMs derived from the NVSS
\citep{taylor2009}. RM SFs are presented as a function of direction
for 80\% of the sky. We find that:

1. Large-scale structures visible in RM amplitude and sign appear
related to structures in the Local ISM, such as radio Loops I and II,
the Gum nebula and some other \HII\ regions. We associate region C$'$
(Table~\ref{SFregions-tab}) with a soft X-ray shadow also visible in
\HI\ at $V_{\rm LSR}\approx -2\ \kms$.

2. SFs near the Galactic poles indicate the presence
of a Faraday screen at both Galactic poles. The variance in RM from
foregound ISM is $\sigma^2_{\rm ISM} = 202\ \radmsq$ and the NGP, and
$\sigma^2_{\rm ISM} = 419\ \radmsq$ at the SGP (120 $\radmsq$ and 307
$\radmsq$ respectively when using
Figure~\ref{SF_poles_comp-fig}). This difference is attributed to a
local asymmetry of electron density and/or properties of the magnetic
field with respect to the Galactic plane.

3. The distribution of RM variance over the sky shows a systematic
trend with Galactic longitude such that lines of sight through the
inner Galaxy have higher RM variance than lines of sight through the
outer Galaxy. This suggests that most or all of the line of sight
through the Galaxy contributes to RM variance. The slope of the SF
does not show the same trends, suggesting its origin is related to local
structures that dominate RM variance on larger angular scales.

4. Radio Loop III and the Orion-Eridanus bubble appear to enhance RM
variance between latitudes $30\degr$ and $50\degr$, and a higher SF
slope is found in these areas. Several other curved filamentary
structures are found in direction of the Galactic anti-center
(Figure~\ref{conditional_RM-fig}).

5. The steepest SFs ($\alpha_2 \approx 1.3$) are
found in the direction of the Gum nebula and the $\zeta$ Oph \HII\
regions, but the steep slope may be a result of a localized enhancement
of the mean RM by these two \HII\ regions.

6. RM variance traced by compact extragalactic radio sources can
account for depolarization of diffuse Galactic emission given the
uncertainties, but the over-all variance is on the low side allowed by
the errors.  Differential Faraday rotation along the line of sight
may increase depolarization as well.

\begin{acknowledgements}

JMS thanks dr. Vadim Uritsky for providing a set of images with different
SF parameters as part of a blind test of the SF analysis. This
research was supported by a Discovery Grant of the Natural Sciences
and Engineering Research Council of Canada to Jeroen Stil.

\end{acknowledgements}

{}

\end{document}